\documentclass[12pt,preprint]{aastex}

\shorttitle{Unpredictability}

\shortauthors{Hudson}

\begin{document}

\title{The Unpredictability of  the Most Energetic Solar Events}

\author{Hugh S. Hudson}
\affil{Space Sciences Lab, UC Berkeley}

\begin{abstract}
Observations over the past two solar cycles show a highly irregular pattern
of occurrence for major solar flares, $\gamma$-ray events, and solar 
energetic particle (SEP) fluences.  
Such phenomena do not appear to follow the direct indices of solar magnetic 
activity, such as the sunspot number.  
I show that this results from non-Poisson occurrence for the most energetic events.  
This Letter also points out a particularly 
striking example of this irregularity in a comparison between the declining 
phases of the recent two solar cycles (1993-1995 and 2004-2006, respectively) 
and traces it through the radiated energies of the flares, the associated SEP 
fluences, and the sunspot areas.  
These factors suggest that processes in 
the solar interior involved with the supply of magnetic flux up to the 
surface of the Sun have strong correlations in space and time, leading 
to a complex occurrence pattern that is presently unpredictable on time 
scales longer than active-region lifetimes (weeks) and not correlated well 
with the solar cycle itself.
\end{abstract}


\section{Introduction}

It has long been known anecdotally that highly energetic solar events do not strictly follow the solar sunspot cycle \citep[e.g.,][]{1987SoPh..109..119G}.
The fact that we only have a few cycles of modern data has made it difficult to describe this
discrepancy quantitatively, especially in view of the small numbers of the most energetic events.
The fossil records typically do not have enough time resolution for to overcome these
problems \citep[but see][]{2001JGR...10621585M}.
These most energetic events include some of the most geoeffective ones, so there we have a clear practical reason for studying their occurrence patterns -- we would like to predict the occurrence of a major event.

The most energetic events also represent the extreme limit of the mechanism that stores energy in the solar corona.
In the consensus view magnetic energy builds up gradually in the corona as a result of stresses imposed from below.
The stressed coronal field then relaxes, by unknown processes, to produce a flare and/or coronal mass ejection (CME).
The energy appears to arrive in the corona as the result of  buoyant motions of current-carrying flux systems \citep[e.g.,][]{2007ApJ...655L.117S}  rather than by the twisting of the coronal field by photospheric surface flows as often assumed in numerical simulations.
The patterns therefore reflect the persistence of the flux-emergence process, which is known to display coherence in both space and time \citep[e.g.,][]{2005A&A...438..349K}, and ultimately must be attributed to the solar dynamo and other processes in the solar interior \citep[e.g.,][]{1998SoPh..181....1R}.

Flare occurrence apparently follows a nonstationary Poisson distribution with time-varying mean rates \citep{1994PhDT........51B,2000ApJ...536L.109W,2001JGR...10629951M} and a clearly power-law dependence on event ``size,'' where this conceptually reflects total event energy but in practice often refers to an observational parameter such as peak X-ray luminosity \citep[e.g.,][]{1971SoPh...16..152D,1991SoPh..133..357H}.
Many studies have shown that flare occurrence follows a flat power-law relationship, $d(logN)/d(logE) = -\alpha$, with $\alpha < 2$.
There are suggested weak dependences of the exponent on the phase in the solar cycle 
\citep{1993ApJ...404..805B,2002SoPh..211..255W} by active region \citep{1997ApJ...475..338K}, and for from star to star \citep[e.g.,][]{1989SoPh..121..375S}.
Such a flat distribution requires a high-energy cutoff to conserve energy, but there is no clear evidence
for such a cutoff yet.

The more energetic the flare, the more likely the occurrence of a CME, although in a few cases an X-class flare will not have a CME association \citep[e.g.,][]{1995ApJ...440..386D}.
For weaker flares, associated CMEs occur much less frequently \citep[e.g.,][]{2006ApJ...650L.143Y}.
The CME distribution must therefore deviate from the flare power law at low event energies, possibly not following a power law at all \citep{1993SoPh..148..359J}.
Interestingly, solar energetic particle fluences do follow a power law, but a significantly flatter one than the flares \citep{1975SoPh...41..189V,1996SoPh..165..337G}; see also \cite{1978SoPh...57..237H}.
The occurrence of solar energetic particles (SEPs) might otherwise be expected to reflect the CME distribution, because CME-driven shocks are known to accelerate SEPs \citep[e.g.,][]{1999SSRv...90..413R,2004ApJ...605..902C}.

In this Letter we report a large specific variation in X-class flare occurrence rate that we trace through similar patterns in SEP fluences and in sunspot areas.
This juxtaposition is consistent with the interpretation of flare occurrence with Biesecker's variable-rate Poisson process, although the small numbers of the most energetic flares means that this interpretation is only weakly grounded in this context.
We instead suggest an origin in correlations of solar interior magnetism on time scales longer than about one rotation period, whose existence will strongly limit flare prediction on these time scales until the interior dynamics is better understood.

\section{X-class Flares}

An X-class flare corresponds to a peak flux of 10$^{-3}$~W/m$^{2}$ in the GOES 
standard 2-8\AA~passband.
Such events lie at the upper end of the occurrence energy distribution function of all flares, and may differ in their temporal occurrence because of the requirement for an upper energy cutoff -- because of this, one cannot assume that the energy distribution continues to have the
same power-law form as the flaring rate changes.
Their small numbers (about 125 in the past solar cycle, from 1996 through 2006) make statistical analyses difficult, and in fact the more energetic of these events may saturate the detectors, which tends to diminish the quality of the statistics.

\begin{figure}
\plotone{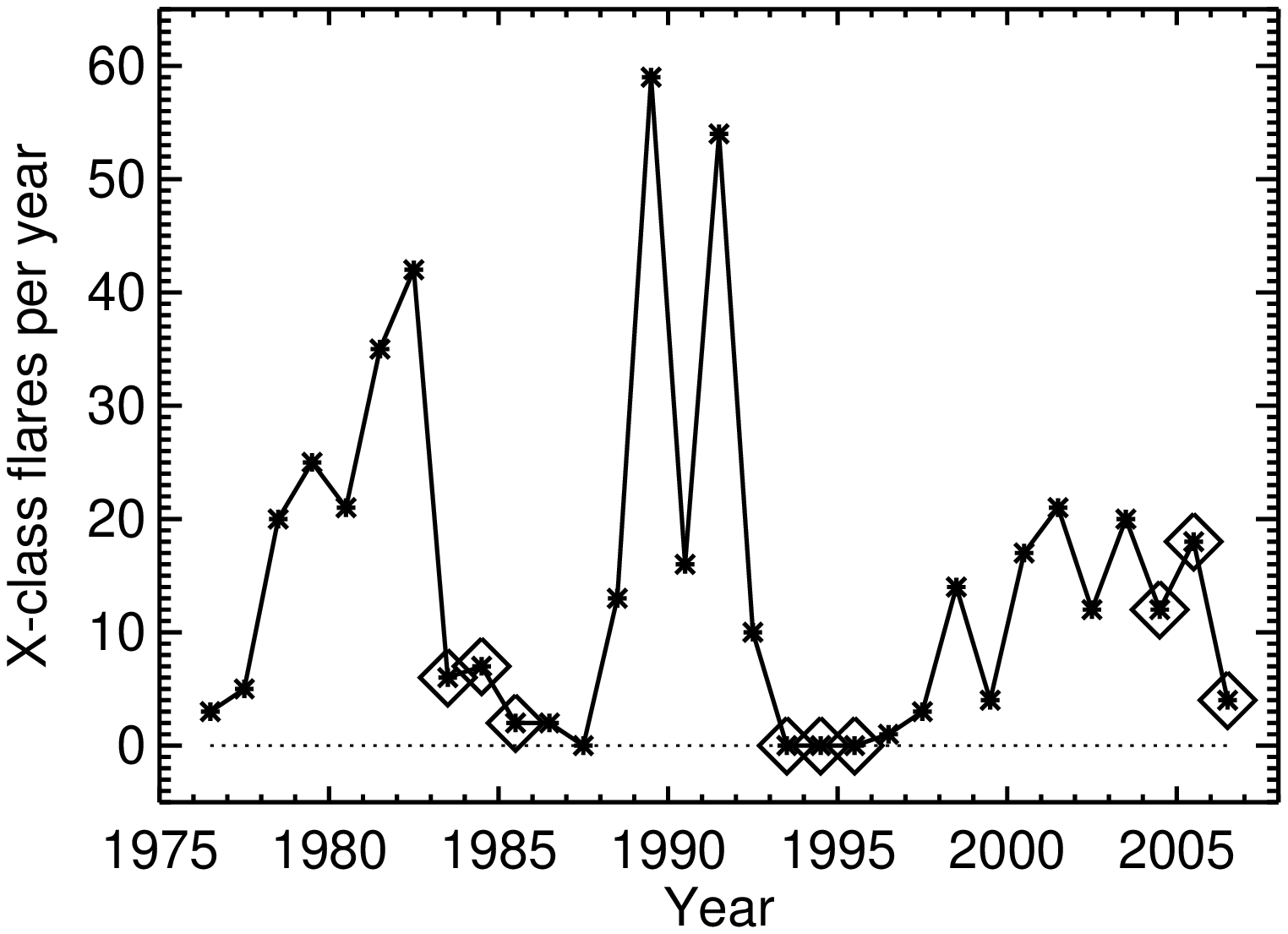}
\caption{X-class flare numbers by year from 1~September 1975 through 31~January 2007.
Points shown as diamonds are the years of the solar-cycle declining phases, defined here as 1983-1985, 1993-1995, and 2004-2006. 
The corresponding numbers of X-class flares are~15, zero, and~34 respectively.}
\label{fig:decay} 
\end{figure}

The declining phases of the past two solar cycles have shown a striking discrepancy in the occurrence
of X-class flares.
This got attention because of the RHESSI observations of $\gamma$-ray flares in 2003-2005
\citep[e.g.,][]{2006SPD....37.0128S}; such events typically correspond to the X-class flares, and 
RHESSI observed several remarkable examples \citep[e.g.,][]{2004ApJ...615L.169S}  in its inaugural years 2002 and 2003.
The expectation for the years 2004-2006, if based on the previous-cycle years of approximately 1993-1995, would have been {\it zero} further events -- not a single X-class flare occurred during these three late years of the previous cycle, although one old-cycle event did occur in 1996 \citep{1998Natur.393..317K,1998opaf.conf..237H}.
To our surprise as many as 34~X-class flares occurred over 2004-2006, though not all observable as $\gamma$-ray events from RHESSI because of its orbital eclipse cycle.
See Figure~\ref{fig:decay} for the data, all of which were obtained from Web resources maintained by NOAA.\footnote{http://www.ngdc.noaa.gov/stp/SOLAR}

Figure~\ref{fig:decay} shows three cycles of X-class flare occurrence, highlighting the discrepant behavior in the decaying phases of Cycles~21, 22~and~23.
The difference in occurrence of energetic events between the latter two epochs is highly significant; for a guide to significance we can use a Poisson distribution based on the number of unique active regions in the years 2004-2006 (11~unique regions, for an average of about 3~X-class flares per region).
Computing the Poisson probability of one event  in the earlier epoch (the 1996 flare) relative to the number of unique regions of the later epoch, we find a likelihood of $<$0.02\%.
This conservatively confirms the obvious inference from the Figure, namely that the X-class event numbers are highly discrepant and that the occurrence of such major energetic events has shown much greater variation than the sunspot number itself.
Cycle~21, on the other hand, showed an intermediate number of events (15~X-class flares, from 9~unique regions) and does not appear discrepant.

\section{Solar Energetic Particles}

\begin{figure}
\plotone{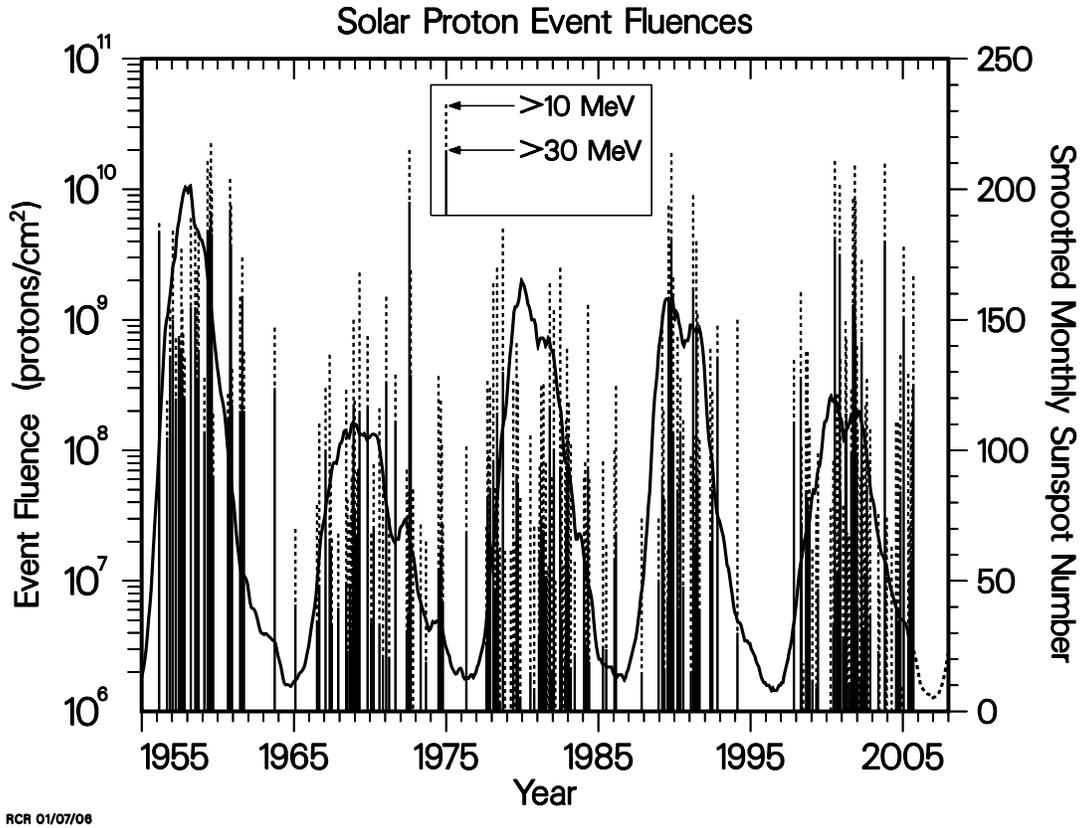}
\caption{Solar energetic particle (SEP) event occurrences for 1954-present (excluding the events of 2006 December), shown as dashed vertical lines for $>$10~MeV threshold and solid vertical lines for $>$30~MeV \citep[from][]{2006LPI....37.1419R}.
The background curve is the sunspot number in monthly bins.
Note the large fluences around 2005, and the negligible fluences one cycle earlier around 1994.
}.
\label{fig:reedy} 
\end{figure}

The striking difference shown by the X-class flare occurrence between the past two cycle declining phases also shows up strongly in the SEP fluences  \citep[Figure~\ref{fig:reedy}, from][]{2006LPI....37.1419R}.
This would be expected because of the strong correlation between X-class flare occurrence and
CME occurrence, as documented recently by \cite{2006ApJ...650L.143Y}.
The declining phases of the two recent cycles, comparing (for example) 1994 with 2005 in Figure~\ref{fig:reedy}, clearly differ significantly.

The identification of flare activity with SEP fluxes might seem inconsistent with the theory of particle acceleration by CME-driven shocks, rather than flares {\it per se} \citep[e.g.,][]{1999SSRv...90..413R,2004ApJ...605..902C}, and frequent assertions of the independence of CME and flare occurrence.
This becomes understandable from the work of \cite{2006ApJ...650L.143Y}, who
confirm the well-known strong association of CMEs with the most energetic flares.
The discrepancy in the numbers of the most energetic events between the two recent cycle declining phases can thus be traced in flare, CME, and SEP occurrence patterns.
We discuss the significance of this finding in Section~\ref{sec:disc} but first investigate whether or not this occurrence discrepancy can also be detected in sunspot area statistics.

\section{Sunspot areas}

\begin{figure}
\plotone{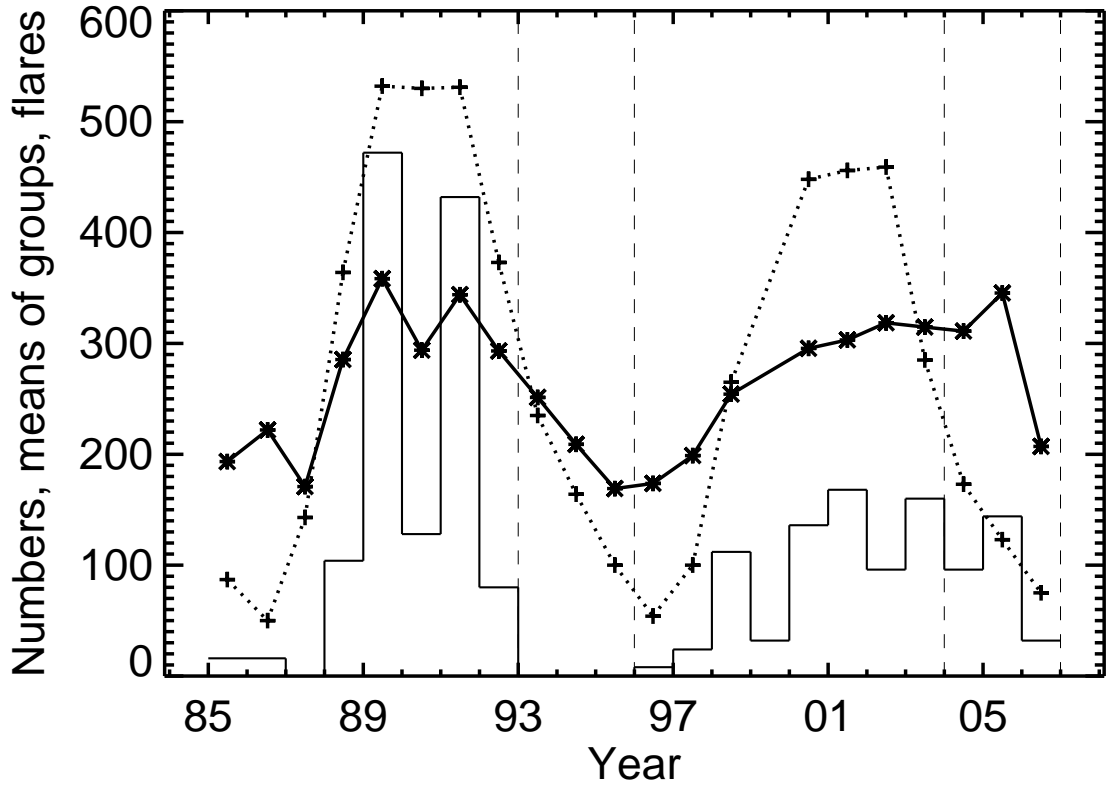}
\caption{Sunspot and flare behavior during Cycles 22 and 23.
Dotted line, the annual numbers of sunspot groups; solid line, 2~$\times$~the mean peak areas of the groups (see text). 
Histogram, the numbers of X-class flares~$\times$~8.
The vertical dashed lines mark the two declining-phase epochs studied in this paper.
Data from the SOON network via NOAA.
}.
\label{fig:spots} 
\end{figure}

The plot in Figure~\ref{fig:spots} shows data obtained from the tabulations of sunspot group area by the SOON$^{1}$ stations.
A large fraction of the tabulated data have been used, typically from three or more stations for each day, but with rejection of a small number of outliers and also the measurements with quality values below~3 (the range is 1-5; see the NOAA Web site for details of the SOON sunspot data).
The solid line in the plot shows the mean of the maxima of the daily areas for individual groups, in millionths of the hemisphere (the customary unit).
This shows a time variation significantly distinct from that of the number of groups (dotted line)
which roughly tracks the sunspot number.
The larger values of mean areas during the decay phase of Cycle~23 (2004-2006) shows that the distribution function of sunspot group areas favored larger spots than during the corresponding interval in Cycle~22 (1993-1995).
This asymmetry coincides with the asymmetry noted above in X-class flare occurrence and in SEP production.

\section{Discussion}\label{sec:disc}

Major energetic solar events do not closely track the solar cycle as a source of the slow variation under the dominant Poisson statistics.
Indeed, the ``Bayesian blocks'' of \citet{2000ApJ...536L.109W} or the time scales
for Poisson behavior obtained by other methods \citep[e.g.,][]{2002SoPh..209..171G}
are considerably shorter than the mean waiting times for X-class events (on the
order of one event per month over 1996-2006).
We conclude that other physics dictates the occurrence patterns of the most energetic events,
for which at most a few may occur in a given active region.
The underlying cause of the Poisson behavior for the less energetic events should be found
in the physics of energy buildup and release in the corona.
The occurrence of the most energetic events presumably has more to do with the
broad-band coherence of solar magnetic activity on large scales in both space and time, as discussed by \cite{2005A&A...438..349K} in terms of ``intermittent oscillations''  revealed by spherical-harmonic expansions of synoptic magnetogram data.
Examples of broad-band correlations would include the butterfly diagram and the presence
of  ``active longitudes'' where active regions may occur repeatedly.
We can also note the remarkable eruption of three distinct active regions in October 2003, each producing X-class flares, and with distinct active regions in both hemispheres.
Such a sudden and widespread surge of activity is certainly remarkable, even though noted here only {\it a posteriori}.

Magnetic flux emergence leads directly to flare activity \citep[e.g.][]{2007ApJ...655L.117S},
and the occurrence of multiple major flares in a given active region therefore points to a persistence in the pattern of flux emergence.
This persistence seems to be required to explain the occurrence of homologous flares, since we believe that extracting the energy from stressed coronal magnetic fields requires their irreversible restructuring, for example by magnetic reconnection.
\cite{2001GeoRL..28.3801N} show that this persistence can result in homologous CMEs
in association with impulsive X-class flares.
For reasons currently unknown, the strongest flux emergence, leading to the most energetic solar events, does not follow the relatively smooth pattern of flux emergence that defines the solar cycle and the occurrence patterns of less-energetic events.

The striking variability in the occurrence of energetic events described in this paper might correspond to a modulation of the event rate near the upper limit on flare energy.
Such a cutoff is required by the non-convergence of the flat occurrence power law of solar flares.
The existence of a cutoff in particle fluences is already well-established from the fossil records, which have the advantage of extending over longer periods of time and thus of capturing the rarer extremely energetic events.
The $^{14}$C record suggests a maximum SEP fluence of some 10$^{10}$ ~protons~cm$^{-2}$ \citep{1980asfr.symp...69L} and fossil cosmic-ray records over longer time scales agree well with this limit \citep{1996ASPC...95..429R}.
\cite{2001JGR...10621585M} set the cutoff at about 6~$\times$~10$^{9}$~protons~cm$^{-2}$ (omnidirectional fluence) at $>$30~MeV based upon nitrate concentrations in Greenland ice cores.
This proxy has the advantage that it overlaps the historical record.

The SEP cutoff fluence corresponds roughly to the largest X-ray flare fluxes, of class~X10 \citep{1980asfr.symp...69L}.
Observing an analogous cutoff in the X-ray fluxes (or other measures of flare energy) is difficult, however, both because of the rarity of the most energetic events and also because they tend to cause detector problems that make it difficult to obtain precise photometry (the GOES\footnote{Geostationary Operational Environmental Satellite} photometers themselves saturate at about this level).
Such a cutoff in X-ray flare statistics, which best reflect total flare energy, has not yet been 
reported.
\cite{2002ApJ...570..423N} actually do observe an upper cutoff in radio burst magnitudes,
in a comprehensive study, but they also note calibration difficulties and other factors that may contribute to this.
The SEP fluxes have a ``streaming flux limit''  \citep[e.g.,][]{1999SSRv...90..413R}, so the agreement of the SEP cutoff with the presently-observed maximum in the GOES event energies may be fortuitous.

\begin{figure}
\plotone{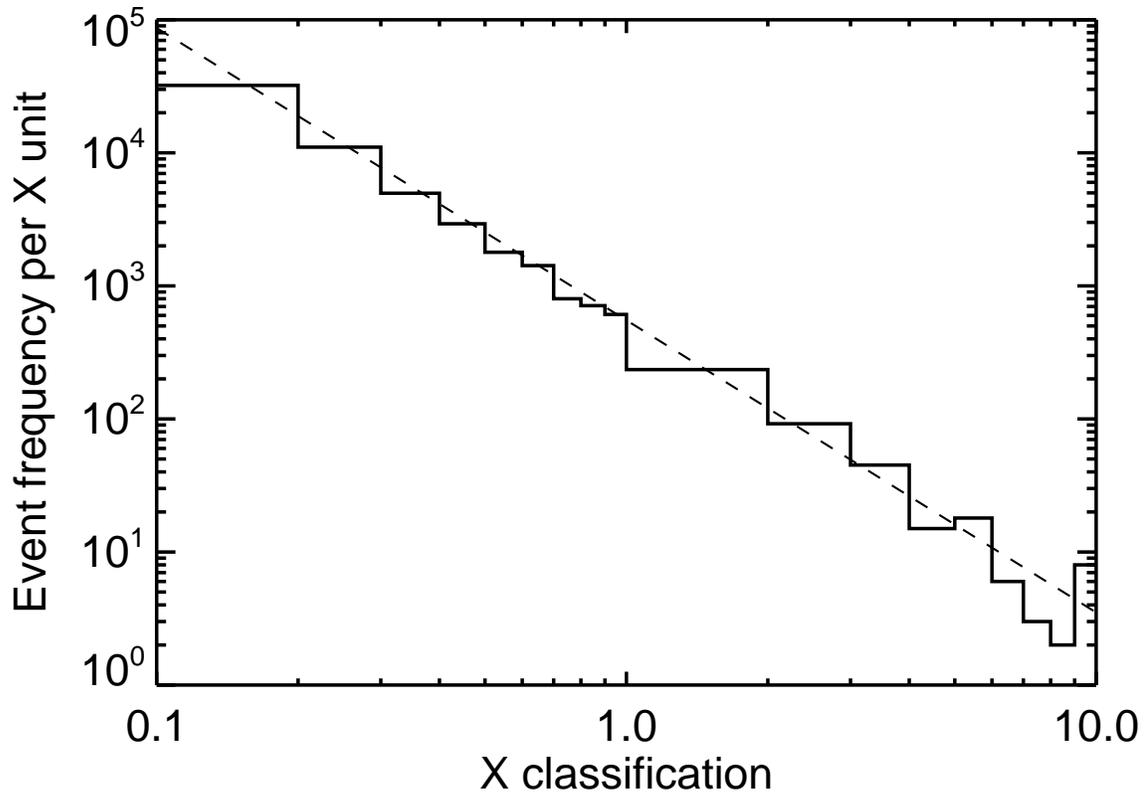}
\caption{Distribution of GOES~1-8\AA~peak fluxes for the interval September, 1975, through January, 2007, for the M~and X-class events (discarding the 22 ``super-flare'' occurrences above X10).
The dashed lines shows a fit using the maximum likelihood method of \cite{1970ApJ...162..405C}, which does not require binning.
The binning shown is 0.1~X~units for the M~flares, and 1~X~unit for the X~flares (where X1~ corresponds to 10$^{-4}$~W/m$^{2}$ peak soft X-ray flux).
This fit predicts the observed number of super-flares within errors, giving a  lower limit on the 
break energy.
}.
\label{fig:goes} 
\end{figure}

Does any index of flare magnitude show a similar high-energy limit?
The soft X-ray photometry from GOES provides the most stable long-term database of flare
magnitudes, and we have analyzed it to answer this question.
Figure~\ref{fig:goes} shows the distribution of M-~and X-class flares for the period from September, 1975, through January, 2007. 
This consists of 5,637~M~events, 424~X~events, and 22~``super'' events above~X10
(numbers inclusive of M1.0, X1.0, and X10.0).
We do not show the super events in the Figure because of distortion due to saturation.
The maximum-likelihood method of \cite{1970ApJ...162..405C}, independent of
binning, gives a fit over the M-X range of $dn/dS\ =\ 5520\ \times S^{-2.193\pm0.015}$ events per unit
X-class interval, the differential distribution.
This distribution predicts 24.6 super-events, whereas 22 were actually observed.
Within errors, there is thus no downward break.
The fit over the M-X range given here is slightly steeper than expected, probably because of the
lack of background subtraction in the reported event magnitudes.
The flare energy upper limit must therefore be significantly above X10 -- as noted by \citet{2000ApJ...529.1026S}, solar super-events, were any to have occurred, ought to have been 
detected by solar astronomers within the historical era.

Resolving this question -- at what point does the flare energy distribution steepen? -- would provide a important clue for students of the generation of solar magnetic flux and its
delivery to the photosphere.
\cite{1997ApJ...475..338K} interestingly suggest that a cutoff may be observable directly in event distributions for smaller active regions, at lower event energies. 
Thus the hypothetical cutoff in X-ray flare magnitudes might reflect the downturn in active-region areas expected from the log-normal distribution noted for sunspot areas \citep{1988ApJ...327..451B}.
The result regarding mean areas (Figure~\ref{fig:spots}) conflicts with the stability of the spot area distribution noted by Bogdan et al., but this may reflect the differing time scales studied.
The existence of the needed cutoff in the distribution has been anticipated by \citet{1975ApJ...200..641M}, who suggested relating the maximum energy of a stellar flare
with the scale lengths present in the convection zone of the star.

\section{Conclusions}

We have shown, based on the decay phases of Solar Cycles~22 and~23, an unexpected example of large-amplitude variations in the occurrence of the most energetic solar events.
We could also trace this pattern in SEP fluxes and in sunspot group areas.
These most energetic events (GOES~X1 or greater) do not follow the usual Poisson statistics with mean
rates that govern lesser flares with shorter waiting times.
The waiting times for the most energetic events indeed often exceed the active-region lifetimes, or the solar rotation period.
Their statistics therefore reflect physics unrelated to coronal energy buildup and the mean flaring 
rate for a given active region.
We suggest that solar interior dynamics dictates the pattern of occurrence of the most energetic events,
rather than the coronal development.

This dramatic variability reduces the predictability of major hazards in space \citep[e.g.,][]{2007P&SS...55..517S}, since it is clear that a variable-rate Poisson distribution following the solar cycle as defined by a smooth sunspot number will not suffice.
Worse yet, the flatness of the particle fluence distribution -- which has an index of 1.2-1.4 \citep{1975SoPh...41..189V,1996SoPh..165..337G}, flatter still than the flare energy distribution at about~1.8 \citep[e.g.,][]{1991SoPh..133..357H} -- means that individual events will dominate the total X-ray and $\gamma$-ray fluences.
At present such events are basically unpredictable on time scales longer than a few days.

\begin{acknowledgements}
This work was supported by NASA NAG5-12878.
I especially thank Bob Lin, Bob Reedy, and Albert Shih for help during the preparation of this paper.
I also thank Ed Cliver for a reading of the preliminary version and Mike Wheatland for correspondence.
\end{acknowledgements}

\bibliographystyle{apj}
\bibliography{xclass}



\clearpage

\clearpage

\clearpage

\end{document}